# WHICH STARS CAN SEE EARTH AS A TRANSITING EXOPLANET?

L. Kaltenegger[1,21*], J.Pepper[3]
[1] Carl Sagan Institute, Cornell University, Space Science Institute 302, 14850 Ithaca, NY, USA,
[2] Astronomy Department, Cornell University, Space Science Institute 302, 14850 Ithaca, NY, USA
[3] Lehigh University, Physics Department, Bethlehem, PA, 18015, USA

**ABSTRACT**

Transit observations have found the majority of exoplanets to date. Spectroscopic observations of transits and eclipses are the most commonly used tool to characterize exoplanet atmospheres and will be used in the search for life. However, an exoplanet's orbit must be aligned with our line of sight to observe a transit. Here we ask, from which stellar vantage points would a distant observer be able to search for life on Earth in the same way?

We use the TESS Input Catalog and data from Gaia DR2 to identify the closest stars that could see Earth as a transiting exoplanet: We identify 1,004 Main Sequence stars within 100 parsecs, of which 508 guarantee a minimum 10-hour long observation of Earth's transit. Our star list consists of about 77% M-type, 12% K-type, 6% G-type, 4% F-type stars, and 1% A-type stars close to the ecliptic. SETI searches like the Breakthrough Listen Initiative are already focusing on this part of the sky. Our catalog now provides a target list for this search. As part of the extended mission, NASA's TESS will also search for transiting planets in the ecliptic to find planets that could detect life on our transiting Earth as well.

## INTRODUCTION

More than 3,000 transiting exoplanets have been detected to date (exoplanets.nasa.gov, July 2020) with dozens of terrestrial planets orbiting in the temperate Habitable Zone of their stars (e.g. Kane et al 2016, Johns et al. 2018; Berger et al. 2018). NASA's Transiting Exoplanet Survey Satellite (TESS) mission (Ricker et al. 2016) has already searched about 74% of the sky in its two-year primary mission for transiting extrasolar planets, including potentially habitable worlds orbiting the closest and brightest stars. The TESS Habitable Zone Star Catalog (Kaltenegger et al 2019) derived the list of stars where TESS can detect transiting Earth-sized planets orbiting in the Habitable Zone of their host star.

Here, we reverse the viewpoint and ask from which systems other observers could see Earth as a transiting planet. Signs of a biosphere in the atmosphere of transiting Earth, such as the combination of oxygen or ozone with the reducing gas methane, could have been detected for about 2 billion years in Earth's history (Kaltenegger et al. 2020). In addition to direct imaging spectroscopy, transit observations are a key to characterizing

---

[1*] lkaltenegger@astro.cornell.edu

inhabited extrasolar planets (see e.g. reviews by Kaltenegger 2017, Fuji et al 2018).

We aim to identify all nearby stars of interest which could see Earth as a transiting exoplanet. This topic has been explored by several teams to identify a priority space for SETI searches (see e.g. Filippova & Strelnitskij 1988, Castellano et al. 2004, Shostak & Villard 2004; Conn Henry et al.2008, Nussinov 2009, Heller & Pudritz 2016, Wells et al. 2018, Sheik et al. 2020) because observing Earth as a transiting planet would classify it as a living planet and thus as an interesting target for deliberate broadcasts.

Previous research has described the Earth Transit Zone (ETZ) as the region from which the Earth could be seen transiting the Sun, which is a thin strip around the ecliptic as projected onto the sky with a width of 0.528° (Heller & Pudritz 2016). Heller & Pudritz (2016) introduced the concept of the restricted Earth Transit Zone (rETZ) as the region which sees Earth transit for more than 10 hours (with a transiting impact parameter $b \leq 0.5$), and thus has a smaller width of 0.262°. The impact parameter denotes the distance of the planet's center from the star's center in units of stellar radii. The Earth's transit duration is 12.6 hrs for an equatorial transit (b = 0) and 10.9 hours for an impact parameter of b = 0.5.

Heller & Pudritz (2016) identified 82 stars within 1,000 pc in the restricted ETZ based on Hipparcos data, with 22 of those stars within 100 pc. The authors extrapolated these findings to estimate that about 500 stars should exist in that region.

Wells et al. (2018) discern 1,022 G and K dwarf stars with V-mag < 13 in the ETZ using data from the SIMBAD astronomical database. These initial analyses, however, were limited by the completeness of the datasets available at the time and the difficulty to differentiate between dwarfs and subgiants.

Our set of stars deliberately excludes evolved stars and those with poorly measured stellar parameters. We now have much more precise measurements of the distances to stars from the Gaia mission Data Release 2 (DR2; Gaia Collaboration et al. 2018). A particular advantage provided by DR2 is the measurement of parallactic distances to fainter, late-type dwarf stars.

Not only do the DR2 data allow us to more carefully situate low-mass stars according to stellar distance, the DR2 distances, when combined with apparent brightness measurements and spectral energy distributions, yield empirical radius measurements of stars, enabling us to differentiate between dwarfs and subgiants. While evolved stars might be able to show atmospheric signs of biota for a specific time in their evolution (see e.g. Agol et al 2014, Loeb & Moaz 2013, Kozakis & Kaltenegger 2018, Kozakis et al. 2020, Kozakis & Kaltenegger 2020), here we focus on Main Sequence stars, which are the main targets in our search for life on exoplanets.

The first observations of the Earth transit zone have recently been undertaken by the Breakthrough Listen Initiative (Sheik et al 2020). In addition, TESS has entered the extended mission phase, with a plan to observe stars in the ETZ in 2021 or 2022. Therefore, a detailed and updated treatment of the ETZ is timely.

The primary tool we use for this analysis is the TESS Input Catalog version 8 (TIC-8; Stassun et al. 2018, 2019). The TIC is a compiled catalog used by the TESS mission to select target stars to optimize the search for transiting planets. Including DR2, as well as a large set of photometric and spectroscopic catalogs, the TIC provides a comprehensive set of stellar distances, effective temperatures, and approximate luminosity classifications.

Section 2 shows how we identified the closest stars, section 3 identifies known exoplanet hosts in our list and section 4 discusses and summarizes our paper.

## 2. IDENTIFYING THE CLOSEST MAIN SEQUENCE STARS IN POSITIONS TO SEE EARTH AS A TRANSITING EXOPLANET

To identify the stars in the Earth transit zone we use TIC-8 and the accompanying Candidate Target List (CTL-8.01; Stassun et al. 2019). TIC-8 contains roughly 1.7 billion stars, in a magnitude limited sample primarily constrained by the magnitude limits of DR2 and 2MASS. The CTL contains a selection of 9.5 million mostly bright, cool, dwarf stars from the TIC, along with physical parameters calculated based on the empirical properties of the constituent catalogs. The CTL also includes a separate set of cool dwarfs, mostly late K- and M-type dwarfs, assembled from the Cool Dwarf Catalog (Muirhead, et al. 2018). The physical parameters of the stars in the CTL include stellar effective temperatures, masses, radii, and luminosities using a set of empirical relations, all described in Stassun et al. (2019).

There are about 10.3 million TIC objects in the ETZ. In order to select nearby Main Sequence stars with reliably measured stellar parameters, we limit our search to the CTL. Note that the selection of stars for inclusion in the CTL uses a quantity included in the TIC as the Gaia data quality parameter, described in the beginning of sections 2.3 and 3.1 of Stassun et al. (2019). That parameter is based on the uncertainties in the Gaia photometric and astrometric measurements and is not in itself a flag that appears in the Gaia DR2 catalog, but instead is derived from tests introduced in the Gaia mission papers. Specifically, it is described with equations 1 and 2 of the DR2 catalog validation paper (Arenou et al. 2018), which in turn was originally defined in the DR2 HRD paper (Gaia Collaboration 2018), and also in Appendix C of the DR2 astrometric solution paper (Lindegren 2018). After the incorporation of DR2 into the TIC, and the release of TIC-8, that quantity was re-cast in a more limited way in the Gaia dataset, using only the Gaia astrometric errors and referred to as the Renormalised Unit Weight Error (RUWE)[2].

We select from all stars with galactic latitudes in the ETZ, spanning the ecliptic latitudes of -0.264° to +0.264°, from the CTL with distances out to 100pc, and identify 1,004 main sequence stars. For the subsection of the restricted ETZ with ecliptic latitude of -0.132° to +0.132°, we find 508 stars which could see Earth transit the Sun for a minimum of 10 hours (see *https://filtergraph.com/etzstars and carlsaganinstitute.org/data)*.

Figure 1 (top) shows the Hertzsprung Russell Diagram for the selected stars in the ETZ, with effective surface temperatures, $T_{eff}$, between 2820 and 8670 K. Fig.1 (bottom) shows the distribution in ecliptic latitude of the stars versus distance: about 77% M stars, 12% K stars, 6% G stars, 4% F stars and 1% A stars, with the closest star at 8.5pc (see Table 2). Note that the star seen in Fig. 1 to lie slightly off the main sequence with $T_{eff}$ around 4000 K and a luminosity of about 0.01 solar is a M dwarf star located about 0.3 degrees from the Galactic Plane. At that proximity to the plane, the derived stellar parameters can be significantly affected by reddening and extinction (see section 2.3.3 of Stassun et al (2019)).

## 3. KNOWN EXOPLANET HOSTS IN OUR STAR LIST

Two of the stars in our list are known exoplanet hosts identified by NASA's K2 mission (Howell et al 2014), which searched part of the ecliptic for transiting planets (Kruse et al 2019) and has confirmed 410 planets to date. K2-155 (EPIC 210897587 d) at 82 pc (04:21:52.49, +21:21:12.95) hosts three known transiting planets (Hirano et al 2018). K2-240

---

[2] https://gea.esac.esa.int/archive/documentation/GDR2/Gaia_archive/chap_datamodel/sec_dm_main_tables/ssec_dm_ruwe.html

(EPIC 249801827) at 70 pc (15:11:24.0, +17:52:31) hosts two known transiting planets. Note that a third known host star in the ETZ, K2-65 (EPIC 206144956) at 72 pc (22 12 50.82, -10 55 31.35) with one known transiting planet (Mayo et al. 2018) is not in our list because the host is a high proper motion star, and as such has missing data in DR2, no RUWE value and a bad Gaia data quality flag in TIC8.

Estimates of the occurrence rate of Earth-size planets in the temperate Habitable Zone of different host stars range between about 50% to 10% (see e.g. overview of different values derived by different teams in Table 1 in Kaltenegger 2017) depending on the limits chosen for the Habitable Zone and size of the planet. However, that number is also strongly affected by the lack of completeness (not all planets have been detected) and reliability of exoplanet catalogs (see Bryson et al 2020 for details). Without adopting any one occurrence rate, we show as an informative example that an occurrence rate of 10% would lead to about 100 Earth-size planets in the temperate Habitable Zone in our sample of 1,004 stars. Higher or lower occurrence rates increase and decrease that number accordingly.

While no Earth-like planets have been detected around the identified stars in the ETZ yet, we provide the orbital distance in AU (Earth analog orbital separation, aEA) as well as orbital period in days (Earth analog orbital period, PerEA) for a nominal planet which receives similar irradiation as Earth for future searches (see Table 1).

## 4. DISCUSSION & CONCLUSION

We identified the closest stars within 100 parsec – 326 lightyears -- with a vantage point to see our Earth as an exoplanet transiting the Sun. We obtained reliable stellar parameters for these 1,004 main sequence stars through the TESS Input Catalog using high quality GAIA DR2 data flags. Because we restricted our search to stars with reliably measured parallactic distances and derived physical parameters, our sample may well be missing some stars that would otherwise meet the selection criteria. Gaia DR2 has known completeness issues regarding high proper motion stars, and as such we may be missing nearby late-type stars in our set, as in the case of K2-65. Efforts to derive a complete census of nearby (d < 100 pc) stars would be a valuable enhancement of this science.

The majority of our sample, 77%, are cool red M stars, with 12% warmer K stars, 6% G stars like our Sun, 4% hotter F stars, and 1% large, hot A stars. The closest star is at a distance of only 8.5 pc – about 28 light years – from our Sun.

We have no clear answer to how much time life needs to originate and evolve on another planet. We have only one example: On Earth, isotopic data indicate that life started by about 3.8 - 3.5 billion years ago (e.g. Mojzsis & Arrhenius, 1997).

The stars in our sample span a wide range of estimated ages: A typical field G, K, or M dwarf star is several billion years old. However, assessing the age of a specific field star reliably is notoriously difficult (e.g. Soderblom 2010). Hotter stellar types like F and A stars have much shorter lifetimes, making them a lesser priority in the search for advanced complex life, when taking life's evolution on Earth as our basis. Age estimates of the stars in this sample - beyond the statistical estimates for field stars of similar dwarf spectral types - would help to further prioritize targets for the search for advanced life.

SETI searches like the Breakthrough Listen Initiative are already focusing part of their search on this part of the sky: Our catalog now provides a target list of 1,004 stars within 100 pc for their search. This list can also be used to target these stars for detections of new transiting planets by TESS in its extended mission, which is scheduled to start to search for transiting planets in the ecliptic in 2021.

For the closest stars, their high proper motion can move them into and out of the vantage point of seeing our Earth block the light from our Sun in hundreds of years: For example, Teergarden's star - which hosts two known non-transiting Earth-mass planets - will enter the ETZ in 2044 and be able to observe a transiting Earth for more than 450 years (Zechmeister et al. 2019) before leaving the ETZ vantage point.

Thus, the stars which could have seen Earth when life started to evolve are a different set to the ones which can spot signs of life on our planet now compared to those which will see it transit in the far future: Therefore, our list presents a dynamic set of our closest neighbors, which currently occupy a geometric position, where Earth's transit could call their attention.

**ACKNOWLEDGEMENTS:** LK acknowledges support from the Carl Sagan Institute at Cornell and the Breakthrough Foundation. We thank René Heller for very constructive comments. Special Thanks to Shami Chatterjee, Meghan Kennedy and Qing Zhao for offering their porch to use as a secondary outside office to write and to Ryan MacDonald for helpful comments on the draft.

**DATA AVAILABILITY** All data are incorporated into the article and its online supplementary material and are also available at *https://filtergraph.com/etzstars*.

**Table 1:** Closest Stars in the Earth Transit Zone, sorted by distance, full table is available online[a].
a) Legend for Table 1

Units  Label   Explanations
--------------------------------------------------------------------------------
TIC    TESS Input Catalog identifier
mag    Tmag    TESS broad band magnitude
mag    Vmag    V band magnitude
K      Teff    Effective temperature
pc     Dis     Distance
R      Solar radii  Stellar Radius
M      Solar mass   Mstar
L      Solar luminosity Luminosity
deg    ELAT    Ecliptic latitude
deg    GLAT    Galactic latitude
deg    RAdeg   Right Ascension in decimal degrees (J2000)
deg    DEdeg   Declination in decimal degrees (J2000)
---    Gaia    Gaia catalog identifier
---    2MASS   2MASS all sky survey catalog identifier
---    Gaia quality flag
---    in the restricted Earth Transit zone
AU     aEA     Earth Analog orbital separation
d      PerEA   Earth Analog orbital period

*data available online: (https://filtergraph.com/etzstars)*

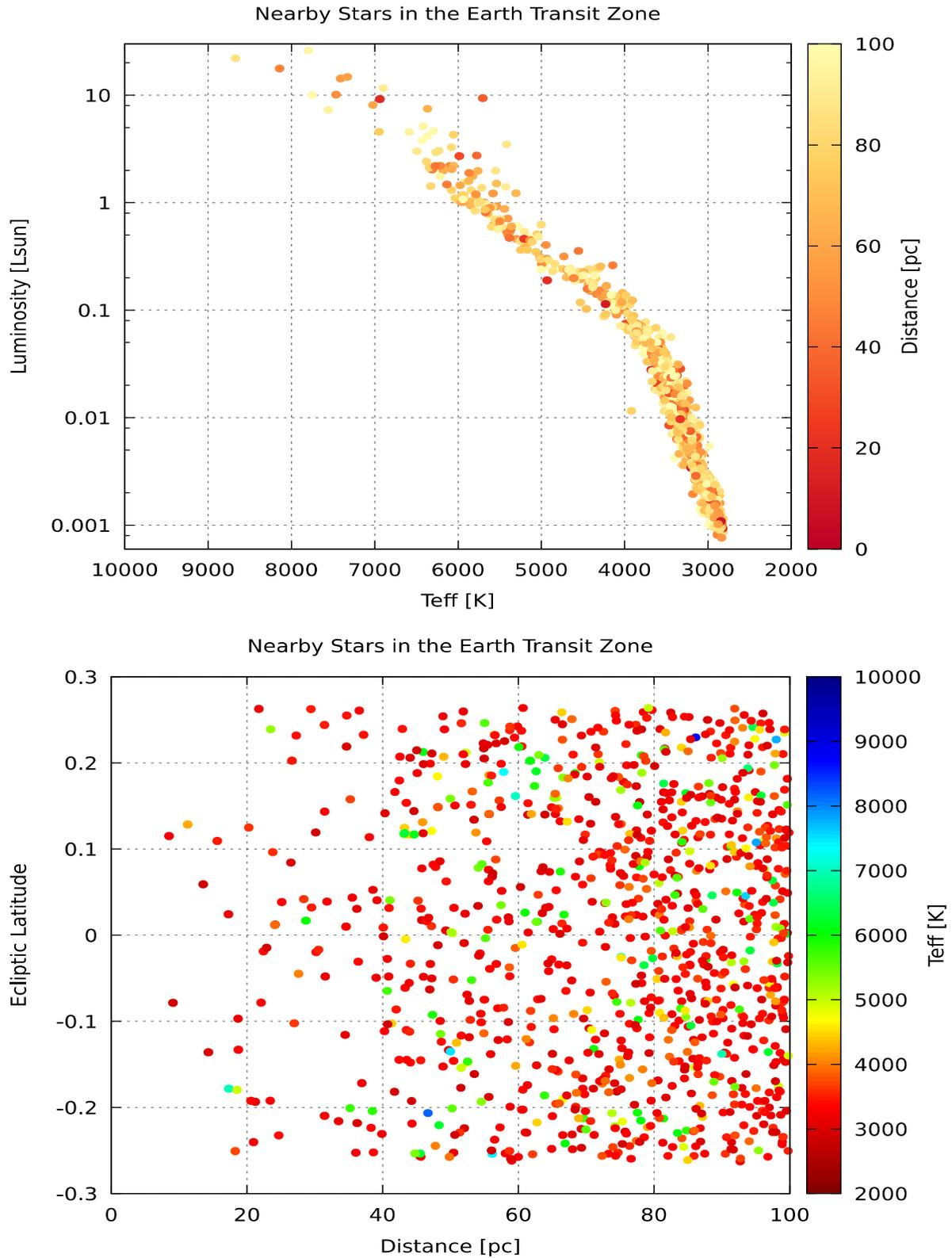

**Figure 1**: Our sample of 1,004 Main Sequence stars within 100 pc, which could see Earth as a transiting exoplanet: (top) Luminosity versus effective temperature, (bottom) Ecliptic latitude versus distance.

**Table 2:** Details on stellar types and closest stars of the 1004 stars in the Earth Transit Zone

| Star type | Number | $T_{eff}$ range [K] | Percentage | Closest (distance, name) |
|---|---|---|---|---|
| M star | 771 | $2820 < T_{eff} < 3840$ | 77 % | 8.5 pc (Ross 64) |
| K star | 124 | $3840 < T_{eff} < 5150$ | 12 % | 11.2 pc (HD 28343) |
| G star | 63 | $5150 < T_{eff} < 5940$ | 6 % | 23.5 pc (HD 65430*) |
| F star | 38 | $5940 < T_{eff} < 7300$ | 4 % | 17.3 pc (Delta Gem*) |
| A star | 8 | $7300 < T_{eff} < 9790$ | 1 % | 46.7 pc (59 Leo) |

*Spectroscopic binary

**Table 3:** Details on stellar types and closest stars of the 509 stars in the restricted Earth Transit Zone

| Star type | Number | $T_{eff}$ range [K] | Percentage | Closest (distance, name) |
|---|---|---|---|---|
| M star | 398 | $2820 < T_{eff} < 3840$ | 78 % | 8.5 pc (Ross 64) |
| K star | 62 | $3840 < T_{eff} < 5150$ | 12 % | 11.2 pc (HD 28343) |
| G star | 29 | $5150 < T_{eff} < 5940$ | 6 % | 40.6 pc (HD 115153) |
| F star | 17 | $5940 < T_{eff} < 7300$ | 3 % | 28.6 pc (29 Ari) |
| A star | 2 | $7300 < T_{eff} < 9790$ | 1 % | 93.5 pc (HD 71988) |